\documentclass[conference]{IEEEtran}
\makeatletter
\def\ps@headings{%
\def\@oddhead{\mbox{}\scriptsize\rightmark \hfil \thepage}%
\def\@evenhead{\scriptsize\thepage \hfil \leftmark\mbox{}}%
\def\@oddfoot{}%
\def\@evenfoot{}}
\makeatother 

\IEEEoverridecommandlockouts
\usepackage{amsfonts}
\usepackage[dvips]{graphicx}
\usepackage{times}
\usepackage{cite}
\usepackage{amsmath}
\usepackage{array}
\usepackage{amssymb}

\newcommand{\bs}{\boldsymbol}

\usepackage{stfloats}
\usepackage{slashbox}
\usepackage{graphicx}
\usepackage{footnote}
\usepackage{booktabs}
\usepackage{array}
\usepackage{algorithmic}
\usepackage{algorithm}
\usepackage{subeqnarray}
\usepackage{cases}
\usepackage{threeparttable}
\usepackage{color}

\begin{document}

\title{Energy Harvesting for Secure OFDMA Systems}

\author{\IEEEauthorblockN{Meng Zhang, Yuan Liu, and Suili Feng}
\IEEEauthorblockA{School of Electronic and Information Engineering\\
South China University of Technology, Guangzhou, 510641, P. R. China\\Email: akjihfdkog@gmail.com, eeyliu@scut.edu.cn, fengsl@scut.edu.cn}

\thanks{This work is supported by the the National Natural Science Foundation of China under grants 61340035 and 61401159, the Science \& Technology Program of Guangzhou under grant 2014J4100246, and the SCUT - UNSW Canberra
Research Collaboration Scheme.
}
}
\maketitle

\vspace{-1.5cm}
\begin{abstract}
   Energy harvesting and  physical-layer security in wireless networks are of great significance. In this paper, we study the simultaneous wireless information and power transfer (SWIPT) in downlink orthogonal frequency-division multiple access (OFDMA) systems, where each user applies power splitting to coordinate the energy harvesting and information decoding processes while secrecy information requirement is guaranteed. The problem is formulated to maximize the aggregate harvested power at the users while satisfying secrecy rate requirements of all users by subcarrier allocation and the optimal power splitting ratio selection. Due to the NP-hardness of the problem, we propose an efficient iterative algorithm. The numerical
results show that the proposed method outperforms conventional methods.

\end{abstract}

\section{Introduction}

   Security is a significant issue in designing and developing wireless systems. Cryptography encryption as a traditional method dominates the upper layers by means of increasing the computational complexity. Security approaches have been introduced in every layer but physical layer in the standard five-layered protocol stack. Thus physical-layer security has been an important complement to the other security approaches.

   A great deal of studies have been devoted to information-theoretic physical-layer security\cite{Jorswieck,Li,Goel,Zhu,Kwan},
In \cite{Jorswieck,Li}, resource allocation for physical-layer security considerations was studied for multicarrier and multiple input multiple output (MIMO) systems. In \cite{Goel}, artificial noise was introduced in physical-layer security to obtain better performance in symmetric Gaussian interference channel. The authors in \cite{Zhu} studied the cooperative power control by using artificial noise in symmetric Gaussian interference channel. In \cite{Kwan}, the authors studied the energy-efficient for
physical-layer security in multicarrier systems.

   On the other hand, simultaneous wireless information and power transfer (SWIPT) becomes an appealing solution to prolong the lifetime of wireless network nodes. It has drawn a great deal of research interests \cite{Zhang,ZhangHo,LiuZ,LiuPS,ZhouZhang}. Energy-harvesting wireless networks are potentially able to gain energy from the wireless environment. The prior work \cite{Varshney,Grover,Fouladgar} studied the performance of transmitting information and energy simultaneously in the receiver that can decode information and harvest energy for the same received signal, which may not be realizable however. Two practical schemes, so-called time switching and power splitting, were proposed in\cite{ZhangHo,ZhouZhang}, as practical designs. With time switching adopted at a receiver, the received signal is either processed for energy harvesting or for information decoding. When the power splitting applied at the receiver, the received signal can be split into two streams with one stream processed by the power harvester and the other processed by the information receiver. Flat-fading channel variations in SWIPT was studied in \cite{LiuZ} and \cite{LiuPS}, where the receiver performs dynamic time switching or dynamic power splitting in the systems, respectively.

Orthogonal frequency division multiplexing access (OFDMA) gains its popularity and has become a leading technology in future broadband wireless networks,
due to its flexibility in resource allocation and robustness against multipath. It enables efficient transmission of various data traffic by optimizing power, subcarrier, or bit allocation among different users.
%
Plenty of works have taken into account either physical-layer security or SWIPT issue in OFDMA, such as \cite{Wang, Kwan,ZhouZhang,HuangLarsson}. For instance, energy-efficient resource allocation for physical-layer security with multiple antennas and artificial noise was investigated in \cite{Kwan}. Resource allocation with SWIPT in OFDMA with different configurations was considered in \cite{HuangLarsson}.

%
 The combination of SWIPT and private message exchanges between mobiles users and base station (BS) can be promising, since it not only meets mobiles users' secrecy requirements and but also enables users to harvest energy. In an OFDMA system, if a specific subcarrier is unsatisfactory for one user to transmit information, to transfer power instead can achieve a high efficient goal and prolong the lifetime of wireless networks. However, SWIPT based OFDMA system with security considerations has not been considered yet in the literature.
This motivates us to investigate physical-layer security in SWIPT based OFDMA systems.

In \cite{Kwan2}, a case of power splitting applied at each subcarrier was considered.  In practical circuits, power splitting is performed before OFDM demodulation. As a result, power splitting should be considered to perform on each user. However, the complexity of splitting on all subcarriers with the same ratio is much greater than splitting on each subcarrier with dynamic ratio. Furthermore, the Lagrangian function cannot be decomposed at each subcarrier. As a result, the optimal solution for power splitting at each user is difficult to obtain.

In this study, we consider the secrecy-rate required multi-users in downlink OFDMA networks, where the users apply power splitting to coordinate energy harvesting and information decoding processes. Our goal is to find efficient subcarrier allocation policies with power splitting ratio to maximize the aggregate harvested energy of all users satisfying the secrecy rate constraints among all users.

First, we formulate a problem as power splitting applied at each receiver for practical application (P-PA). We next formulate a problem as performance upper bound (P-UB) by assuming that power splitting is able to be applied at each subcarrier. Both problems are formulated as mixed integer programming problems and NP hard. For Problem (P-UB), the optimal solution is obtained by dual method. For Problem (P-PA), we propose an efficient iterative algorithm to find the solution that is very close to the performance
upper bound.
%

%

\section{System Model And Problems Formulation}

In this paper, we consider the downlink of an OFDMA network with one BS, $K$ mobile users, over $N$ subcarriers. Each user communicates with BS and demands a secrecy rate that is no lower than a constant $C_{k}$, for all $1\leq k\leq K$. Here we assume that equal power allocation is performed by the BS over all subcarriers for simplicity. This is reasonable since the gain brought by power adaption is very limited in OFDMA systems \cite{YuanTWC10,YuanTCOM12,YuanTWC13,YuanWCL12,TaoYuanTWC13}.
The receivers are considered to split the received signal into two signal streams, with one stream to the energy receiver and the other one to information receiver. Furthermore, we assume for each user, any other user in the same network is a potential eavesdropper. The BS can obtain full statistical knowledge and instantaneous knowledge of CSI and
each subcarrier is occupied only by one user at each time.
This is reasonable since all users are assumed to be legitimate users and have their own data transmission with the BS.


Let $h_{k,n}$ denote the channel gain of user $k$ on subcarrier $n$, and $\beta_{k,n}$ denote the channel gain of the potential eavesdropper for user $k$ on subcarrier $n$, i.e., $\beta_{k,n}=\max_{k',k'\neq k}h_{k',n}$.
We use a binary assignment variable $x_{k,n}$ to represent the subcarrier allocation, with $x_{k,n}=1$ indicating that subcarrier $n$ is assigned to user $k$ and $x_{k,n}=0$ otherwise. Let $p_{n}$ represent the equal power allocated on subcarrier $n$.

All subcarriers have to be power split from the RF signals. Energy harvesting has to perform in analog domain instead of the digital domain where information of each subcarrier is decoded. As a result, due to the hardware limitation, for each mobile station, the received signal has to be harvested with a same power splitting ratio on all subcarriers.

We first consider the practical scenario, that is the received signal at user $k$ is processed by a power splitter, where a ratio $\rho_{k}$ of power is split to energy receiver and remaining ratio $1-\rho_{k}$ of power is split into the information decoder, with $0\leq \rho_{k} \leq 1$, $\forall k$. Thus, the achievable secrecy rate at subcarrier $n$ when assigned to user $k$ is

\begin{align}\label{eqn:pb2}
r_{k,n}^{s}= & \left[ \log\left ( 1+\frac{\left ( 1-\rho _{k} \right )p_{n}h_{k,n}}{\sigma ^{2}} \right)
-\log\left ( 1+\frac{p_{n}\beta _{k,n}}{\sigma ^{2}} \right ) \right]^{+}\nonumber\\
= &\left [ \log\left ( \frac{\left ( 1-\rho _{k} \right )p_{n}h_{k,n}+\sigma ^{2}}{p_{n}\beta_{k,n}+\sigma ^{2}} \right ) \right ]^{+},
\end{align}
where $[\cdot]^+=\max\{\cdot,0\}$.

Then the secrecy rate of user $k$ is denoted as

\begin{equation}\label{eqn:pb1}
     r_{k}^{s}=\sum_{n=1}^{N}x_{k,n}r_{k,n}^{s}.
\end{equation}

Assuming that the conversion efficiency of the energy harvesting process at each receiver is denotes by  $0<\zeta <1$, the harvested power of user $k$ is thus given by
     \begin{equation}
     E_{k}=\zeta \rho _{k} \sum _{n=1}^{N}p_{n}h_{k,n}.
     \end{equation}

The goal of the considered problem is to find the optimal subcarrier allocation and power splitting ratio in order to maximize the total harvested power (for the purpose of uplink transmission for example) while satisfying the individual secrecy rate requirement for each user. This practical application optimization problem can thus be expressed as
     \begin{eqnarray}
 (P-PA): \max_{\{\bs X, \bs\rho\}}&&\zeta \sum _{k=1}^{K}\rho _{k} \sum _{n=1}^{N}p_{n}h_{k,n}\label{eqn:max}\\
s.t.~&&\sum _{k=1}^{K}x_{k,n}\leq 1, \forall n \label{eqn:conx1}\\
  &&x_{k,n} \in \left \{ 0,1 \right \}, \forall k,n\label{eqn:conx2}\\
  &&0\leq \rho_{k} \leq 1,\forall k\label{eqn:conrho}\\
&&\sum _{n=1}^{N}x_{k,n}r_{k,n}^{s}\geq C_{k}, \forall k\label{eqn:conr} \end{eqnarray}
where $\bs X=\{x_{k,n}\}$, and $\bs \rho=\{\rho_{k}\}$.
The constraints in \eqref{eqn:conx1} and \eqref{eqn:conx2} enforce that each subcarrier can only be used by one user to avoid multi-user interference.

An upper bound for this problem can be obtained by assuming the power splitting is applied in receiver to split power for each subcarrier\footnote{This is because of the improved flexibility of resource allocation. In this case, users can harvest all the power at the  subcarriers that are not assigned to them and thus improve the harvested power, i.e., $\rho_{k,n}=1$ for all $x_{k,n}=0$.}. In this case, $\rho_k$ is extended to $\rho_{k,n}$. Thus, we consider the following optimization problem as

     \begin{eqnarray}
  (P-UB): \max_{\{\bs X, \bs\rho\}}&&\zeta\sum _{k=1}^{K}  \sum _{n=1}^{N}\rho _{k,n}p_{n}h_{k,n}\label{eqn:maxuuper}\\
s.t.~ &&\eqref{eqn:conx1}, \eqref{eqn:conx2}, \eqref{eqn:conr}\nonumber\\
      &&0\leq \rho_{k,n} \leq 1, \forall k,n
     \end{eqnarray}
where
\begin{equation}
r_{k,n}^{s}=\left [ \log\left ( \frac{\left ( 1-\rho _{k,n} \right )p_{n}h_{k,n}+\sigma ^{2}}{p_{n}\beta_{k,n}+\sigma ^{2}} \right ) \right ]^{+}\label{eqn:rrr}.
\end{equation}

\section{Optimal Power Splitting Ratio Selection and Subcarrier Allocation For Practical Application}
The formulated problem (P-PA) is non-convex, finding the optimal solution is usually prohibitively due to the complexity.
However, according to \cite{noncon}, the duality gap becomes zero in multicarrier systems as the number of subcarriers
goes infinity for satisfying time-sharing condition.

We define $ \mathcal{T} $ as all sets of possible $\bs X$ that satisfy \eqref{eqn:conx1} and \eqref{eqn:conx2}, $ \mathcal{R} $ as all sets of possible $\bs \rho$ for given $\bs X$ that satisfy $0\leq\rho_{k}\leq1$ for $x_{k,n}=1$ and $\rho_{k}=1$ for $x_{k,n}=0$.

 The Lagrangian function for this problem is given as follow:
  \begin{align}
 &L(\bs \rho ,\bs X,\bs \mu )\nonumber\\
 &=\zeta \sum_{k=1}^{K}  \rho _{k}\sum _{n=1}^{N} p_{n}h_{k,n}+\sum _{k=1}^{K}\mu _{k}\left (\sum _{n=1}^{N} x_{k,n}r_{k,n}^{s}-C_{k} \right ) \nonumber\\
 &=\sum _{k=1}^{K}\sum _{n=1}^{N}\left (\zeta \rho _{k}p_{n}h_{k,n} +x_{k,n}\mu _{k}r_{k,n}^{s} \right )-\sum _{k=1}^{K}\mu _{k}C_{k},\label{eqn:lagfuc}
\end{align}
where $\bs \mu = \left[\mu_1,\mu_2,...,\mu_k\right]^{T}$ are the Lagrange multipliers. The dual function is then defined as
  \begin{equation}
g(\bs \mu)=\max_{\bs X \in \mathcal{T} ,\bs \rho \in \mathcal{R}(\bs X)}L(\bs \rho ,\bs X,\bs \mu )\label{eqn:dual2}.
\end{equation}

Then the dual problem is thus given by $\min_{\bs \mu \succeq 0}g(\bs \mu)$.

Through simple observation, when considering the maximization problem in \eqref{eqn:dual2}.
The Lagrangian function cannot be decomposed into $N$ subproblems. Because optimal $\rho_k^*$ has to be computed considering all subcarriers that are assigned to the $k$th user, instead of only considering one specific subcarrier.

Thus, we will obtain a suboptimal solution by iteratively optimizing $\bs X$ with a fixed set of $\bs \rho$, and then optimizing $\bs \rho$ with a fixed set of $\bs X$, which is known as block-coordinate descent method \cite{Iteration}.

\subsection{Optimality Condition of Subcarrier Assignment}
First, we notice that the problem in \eqref{eqn:dual2} can be decomposed into $N$ subproblems at each subcarrier in order to solve the $\bs X$ with a given set of $\bs \rho$ as

   \begin{equation}
  \max_{\bs X_n \in \mathcal{T}}L_{n}(\bs X_n)=\zeta p_{n}\sum _{k=1}^{K}\rho _{k}h_{k,n}+\mu _{k}r_{k,n}^{s} , \label{eqn:solvex}\\
   \end{equation}
which can be solved independently.

Note that for the Lagrangian \eqref{eqn:lagfuc}, we have
\begin{equation}
L=\sum_{n=1}^N L_n - \sum_{k=1}^K \mu_k C_k.
\end{equation}

As a result, the optimal ${x_{k,n}}$ can be obtained as
   \begin{eqnarray}\label{eqn:opxpb2}
     x_{k,n}^{*}=\begin{cases} 1, ~{\rm if}~k=k^{*}=\arg \max_{k}\mathcal{X}_{k,n}\\
     0, ~{\rm otherwise},\end{cases}
   \end{eqnarray}
where $\mathcal{X}_{k,n}= \zeta p_{n}\sum_{k=1}^{K}\rho _{k}h_{k,n}+\mu _{k}r_{k,n}^{s} $.
\subsection{Optimal Power Splitting Ratio}

Next, we consider the problem in \eqref{eqn:dual2} with a given set of feasible $\bs X$. The problem thus can be decomposed into $K$ subproblems and each for one user, which can be solved independently. This subproblems at user $k$ is

   \begin{eqnarray}
  \max_{\rho_k \in \mathcal{R}(\bs X)}&&L_{k}(\rho_{k})=\sum _{n=1}^{N}\left ( \zeta \rho _{k}p_{n}h_{k,n}+x_{k,n}\mu _{k}r_{k,n}^{s} \right )\label{eqn:solverho},\nonumber\\
   \end{eqnarray}
where
   \begin{equation}
   L=\sum_{k=1}^K L_k - \sum_{k=1}^K \mu_k C_k.
   \end{equation}

According to Karush-Kuhn-Tucker (KKT) conditions, the optimal $\rho_{k}^*$ has to satisfy the following equation:
   \begin{align}
   &\frac{\partial L_{k}}{\partial \rho_{k}}=\sum_{n=1}^{N} \left [ \zeta p_{n}h_{k,n}-\frac{\mu_{k}x_{k,n} h_{k,n}p_{n}}{\ln 2\left ( h_{k,n}p_{k,n}\left ( 1-\rho _{k} \right)+\sigma ^{2} \right )} \right ]=0.\label{eqn:cannotsolve}
   \end{align}

 According to \eqref{eqn:cannotsolve}, there is no closed-form for the optimal $\rho_{k}^{*}$. However $\frac{\partial L_{k}}{\partial \rho_{k}}$ monotonically decreases as $\rho _{k}$ increase. Thus a bisection search method for optimal $\rho_{k}^*$ is feasible.

With the fixed $\bs \rho$, the optimal $\bs X$ can be obtained by \eqref{eqn:opxpb2}. The optimal value of Problem (P-PA) can be increased by optimizing $\bs X$. Thus the above process can be iterated until the optimal value of problem ceases to improve. In this case, the updated subcarrier allocation variables $\bs X$ remain the same as they are before the update.
Since We notice that the local optimal solution depends on the initial sets of $\bs \rho$, $M$ sets of $\bs \rho$
are initialized to obtain a robust solution. We obtain the optimal $\bs X$ for each initialization step of $\bs \rho$ and update
the $\bs \rho$ by a bisection search so as to obtain the maximal Lagrangian, and the iterative algorithm is applied to obtain a local optimal solution.
Then, the final solution is the one achieves the greatest harvested power.

In general, as the $M$ increases, greater robustness and optimality of the algorithm can be guaranteed. As $M \to \infty$, this
algorithm is considered to perform as well as the optimal algorithm. However, large number of initialization steps increases the
computation complexity, which will not be favorable for practical applications.
\subsection{Subgradient updating}

Finally, in \cite{Boyd}, the dual function in \eqref{eqn:dual2} is always convex. By simultaneously updating $\bs \mu$, we can solve this problem by subgradient method. The dual variables $\bs \mu$ are updated in parallel as follow
   \begin{align}
\mu_k^{(t+1)}=\left[\mu_k^{(t)}+\alpha_k\left ( C_{k}-\sum_{n=1}^{N}x_{k,n}r_{k,n}^{s}  \right)\right]^+,\forall k. \label{eqn:update1}
   \end{align}

   \begin{algorithm}[tb]
\caption{Proposed Iterative Algorithm}
\begin{algorithmic}[1]
\STATE \textbf{initialize} Randomly generate $\bs M$ feasible sets of $\bs X$ as different initialization steps.

\FOR {each set of initialization}
\STATE \textbf{initialize} $\bs {\mu}$.
\REPEAT
\REPEAT
\STATE Compute $r_{k,n}^{s}$ for all $k$, and $n$ according to \eqref{eqn:pb2}.
\STATE Solve assignment variables $\bs X$ according to \eqref{eqn:opxpb2}.
\FOR {each optimal $\rho_{k}$ to achieve the maximal $L_k$}
\STATE \textbf{initialize} $\rho_{k}^{UB}=1$ and $\rho_{k}^{LB}=0$, then $\rho_{k}=\left ( 1/2 \right )\left ( \rho_{k}^{UB}+\rho_{k}^{LB} \right )$.
\REPEAT
\STATE Compute $\frac{\partial L_{k}}{\partial \rho_{k}}$ according to \eqref{eqn:cannotsolve}.
\IF {$\frac{\partial L_{k}}{\partial \rho_{k}}>0$}
\STATE Set $\rho_{k}^{LB}=\rho_{k}$.
\ELSE
\STATE Set $\rho_{k}^{UB}=\rho_{k}$.
\ENDIF
\UNTIL {$\left |\frac{\partial L_{k}}{\partial \rho_{k}} \right|<\varepsilon $.}
\ENDFOR
\UNTIL {$\bs X$ converge.}
\STATE Update $\bs\mu$ by \eqref{eqn:update1} according to ellipsoid method.
\UNTIL {$\bs\mu$ converge.}
\ENDFOR
\STATE Find the one which achieves the maximum harvested power from $M$ solution.

\end{algorithmic}

\end{algorithm}

To summarize, the above iterative algorithm to solve Problem (P-PA) is given in Algorithm 1. For this algorithm, the computational complexity mainly lies in the steps 2)-23). As the $\rho_{k}$ are obtained individually by bisection search, the complexity of steps 11)-20) is $\mathcal{O}(K)$. Hence, we can also obtain the complexity of steps 5)-21) is $\mathcal{O}(K+KN)$. Next, the complexity of subgradient updates is polynomial in $K$ \cite{Boyd}. We have the computational complexity of steps 2)-23) is $\mathcal{O}(K^{q+1}+K^{q+1}N)$, where $q$ is a constant.
Finally, considering further the $M$ initialization steps, the time complexity of this algorithm is $\mathcal{O}(K^{q+1}M+K^{q+1}NM)$.

\section{Performance Upper Bound}

Then we can derive the Lagrangian function of Problem (P-UB) as follows
 \begin{align}
 &L(\bs X,\bs \rho ,\bs \lambda )=\sum_{k=1}^{K} \sum _{n=1}^{N}\zeta \rho _{k,n}p_{n}h_{k,n}\nonumber\\
&+\sum _{k=1}^{K}\lambda _{k}\left (\sum _{n=1}^{N} x_{k,n}r_{k,n}^{s}-C_{k} \right ) \nonumber\\
 &=\sum _{k=1}^{K}\sum _{n=1}^{N}\left (\zeta \rho _{k,n} p_{n} h_{k,n} +x_{k,n}\lambda _{k}r_{k,n}^{s} \right )-\sum _{k=1}^{K}\lambda _{k}C_{k},\label{eqn:Lag}
\end{align}
  where $\bs\lambda =\left [ \lambda _{1},\lambda _{2},...,\lambda _{K} \right ]^{T}$ is the vector of dual variables.

The Lagrangian dual function can be obtained as

 \begin{equation}
  g\left ( \bs\lambda  \right )=\max_{\bs X \in \mathcal{T}, \bs\rho \in  \mathcal{R}(\bs X)}L(\bs X,\bs \rho ),
\end{equation}
and we can obtain the dual problem as $\min_{\bs \lambda \succeq  0}~g\left ( \bs\lambda  \right )$.

The dual function $ g\left ( \bs\lambda  \right )$ can be  decomposed into $N$ subproblems which can be solved independently. The subproblems can be obtained as
   \begin{align}
 \max_{\bs X \in \mathcal{T}, \bs\rho \in  \mathcal{R}(\bs X)} L_{n}(\bs X_{n},\bs \rho_{n})=\zeta p_{n}\sum _{k=1}^{K}  \rho _{k,n}h_{k,n}+\lambda _{k}r_{k,n}^{s} \label{eqn:subp},
   \end{align}
   where
\begin{equation}
 L=\sum_{n=1}^N L_n - \sum_{k=1}^{K}\lambda_k C_k.
\end{equation}

   We first seek for the optimal power splitting ratio of each subcarrier. According the
   Karush-Kuhn-Tucker (KKT) conditions \cite{Boyd}, we take the partial derivation of $L_{n}(\bs X_{n},\bs \rho_{n})$ with respect to $\rho_{k,n}$, and equate it to zero.

   \begin{align}
\frac{\partial L_{n}}{\partial {\rho_{k,n}}}=&\sum_{k=1}^{K} \zeta p_{n}h_{k,n}-\frac{\lambda_{k} h_{k,n}p_{n}}{\ln 2\left ( h_{k,n}p_{k,n}\left ( 1-{\rho _{k,n}} \right)+\sigma ^{2} \right )}=0.
   \end{align}

 Note that $\frac{\partial r_{k,n}^{s}}{\partial \rho_{k,n}}=0$ when $\rho_{k,n} \leq 1-\beta_{k,n}/h_{k,n}$. Thus the above equation is only true when $\rho_{k,n} > 1-\beta_{k,n}/h_{k,n}$.

   When ${\rho_{k,n}} \leq 1-\beta_{k,n}/h_{k,n}$, we have

      \begin{align}
   \frac{\partial L_{n}}{\partial \rho_{k,n}}=&\sum_{k=1}^{K}\zeta p_{n}h_{k,n},
   \end{align}
   which is always greater than zero.

Thus, the optimal solution $\rho_{k,n}^{*}$ for both cases can be readily given by
      \begin{eqnarray}
&&\rho_{k,n}^{*}=\begin{cases} \dot{\rho_{k,n}}, ~{\rm if}~\dot{\rho_{k,n}} \leq 1-\beta_{k,n}/h_{k,n}\\
     1, ~{\rm otherwise},\end{cases}\label{eqn:oprho} \end{eqnarray}
where
      \begin{align}
         &&\dot{\rho_{k,n}}=\left[1-\frac{\lambda _{k}h_{k,n}}{\zeta \ln2 p_{n}\sum_{k=1}^{K}h_{k,n}}+\frac{\sigma ^2}{\ln2 h_{k,n}p_{n}}\right]_0^1\label{eqn:oprho2},
   \end{align}
   and $[\cdot]^b_a=\max\{\min\{\cdot,b\},a\}$.

   Next, substituting \eqref{eqn:oprho} into $L_{n}(\bs X_{n},\bs \rho_{n})$, the optimal subcarrier assignment policy is given by (the details are easy and omitted here).
   \begin{eqnarray}\label{eqn:opx}
     x_{k,n}^{*}=\begin{cases} 1, ~{\rm if}~k=k^{*}=\arg \max_{k}\mathcal{H}_{k,n}\\
     0, ~{\rm otherwise},\end{cases}
   \end{eqnarray}
where $\mathcal{H}_{k,n}= \sum_{k=1}^{K}\zeta \rho _{k,n}^{*}p_{n}h_{k,n}+\lambda _{k}r_{k,n}^{s} $.

   Then, dual variable $\bs\lambda$ can be updated as follow
    \begin{equation}\label{eqn:update}
\lambda_{k}^{(t+1)}=\left[\lambda_{k}^{(t)}+\alpha_k\left ( C_{k}-\sum_{n=1}^{N}x_{k,n}r_{k,n}^{s}  \right)\right]^+, \forall k.
    \end{equation}

The complexity of the dual based algorithm is analyzed as follows.
For each subcarrier, $ \mathcal{O} ({K}) $ computations are needed. Since the calculation is independent at each subcarrier,
the complexity if $\mathcal{O} ({KN})$ for each iteration. Last, considering the the complexity of subgradient updates,
the overall complexity of subgradient method is $\mathcal{O} ({K^{q+1}N})$,
where $q$ is a constant.
Finally, we present the whole algorithm in Algorithm 2.

    \begin{algorithm}[tb]
\caption{Dual-based method Algorithm for Upper Bound}
\begin{algorithmic}[1]
\STATE \textbf{initialize} $\bs\lambda$.
\REPEAT \STATE Compute
$\rho_{k,n}$  according to \eqref{eqn:oprho} and \eqref{eqn:oprho2}, and then $r_{k,n}^{s}$ according to \eqref{eqn:rrr} for all $k$ and $n$.
\STATE Solves assignment variables $x_{k,n}$ according to \eqref{eqn:opx}.
 \STATE
Update $\bs\lambda$ via \eqref{eqn:update} according to ellipsoid method.
 \UNTIL{$\bs\lambda$ converge.}
\end{algorithmic}

\end{algorithm}

\section{Numerical Results}

In this section, we provide some numerical results to evaluate the performance of the proposed iterative algorithm. In the simulation setup, we consider an OFDMA network with $N=128$ and $K=8$ mobile users. The users are Rayleigh distributed between a reference distance $d_0$ and $10$m, where $d_0=1$m and results in $-30$ dB path loss. In addition, the power is allocated to each subcarrier uniformly, which can be also described as $p_{n}=P_t/N$, where $P_t$ is the total transmit power of the BS. Let $E_{sum}$ denote the overall power harvested by all users. For all energy receivers, it is assumed that $\zeta=0.4$. The minimum secrecy rate $\bar{C}$ is assumed for the first 4 of the users, i.e., $C_{1}=C_{2}=C_{3}=C_{4}=\bar{C}$. And for the rest 4 users, they have no requirement of secrecy rate, i.e., $C_{5}=C_{6}=C_{7}=C_{8}=0$. For the information receiver, $\sigma^2=-30$ dBm.
In addition, for the proposed iterative algorithm, we set $M=200$.

To evaluate the schemes, we introduce another two schemes in this simulation as benchmarks. In the first scheme, denoted as fixed power splitting (FPS), power splitting ratio $\rho_{k}=0.5 ,\forall k$, is fixed for all users beforehand while $\bs X$ are computed according to \eqref{eqn:opx}. In the second scheme, the subcarrier assignment is fixed (FSA), while $\rho_{k}$ for each user is optimized by a bisection search as Algorithm 1.

\begin{figure}[t]
\begin{centering}
\includegraphics[scale=.65]{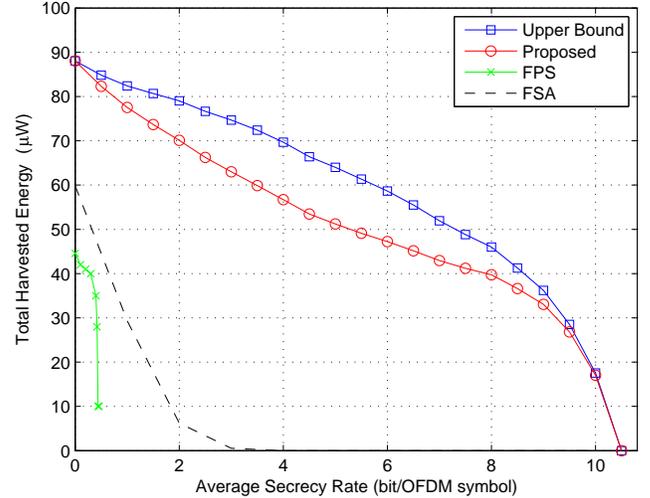}
\vspace{-0.1cm}
 \caption{Achievable $E_{sum}$ versus $\bar{C}$ pair at total transmit power of $15$ dBm.}\label{fig:f1}
\end{centering}
\vspace{-0.3cm}
\end{figure}

 We first demonstrate the pairs of achievable overall harvested power $E_{sum}$ and feasible secrecy rate requirement $\bar{C}$. Fig. \ref{fig:f1} shows the rate pair at fixed total transmit power $P_{t}=15$ dBm. First, it is observed that for all schemes, $E_{sum}$ decreases with the increase of secrecy requirement $\bar{C}$. In addition, for the proposed algorithm and upper bound, $E_{sum}$ falls sharply to zero at around $\bar{C}=10.5$ bit/OFDM symbol.
It is observed that according to the performance of upper bound and the proposed algorithm, power splitting on each user only incurs a little in harvested power when achieving the same $\bar{C}$. On the other hand, the two algorithms earn great advantage over these two benchmarks FPS and FSA. In addition, the maximum feasible points of the FPS and FSA appear at around $\bar{C} = 0.5 $ and 3.8 bit/OFDM symbol respectively, which are much lower than that in the proposed algorithm.

\begin{figure}[t]
\begin{centering}
\includegraphics[scale=.65]{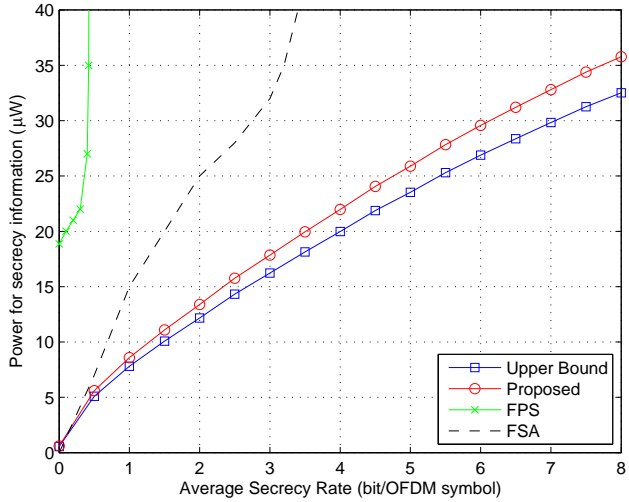}
\vspace{-0.1cm}
 \caption{Total power consumption for all information receivers versus $\bar{C}$ pair at total transmit power of $15$ dBm.} \label{fig:f2}
\end{centering}
\vspace{-0.3cm}
\end{figure}

Fig. \ref{fig:f2} show the overall power consumption for information receivers of all users versus $\bar{C}$. We notice that the proposed algorithm only consumes a little bit more power than the upper bound when achieving the same $\bar{C}$. In addition,
the power consumption for FSA and FPS increases sharply at around $\bar{C} = 0.5 $ and 3.8 bit/OFDM symbol respectively.

\begin{figure}[t]
\begin{centering}
\includegraphics[scale=.65]{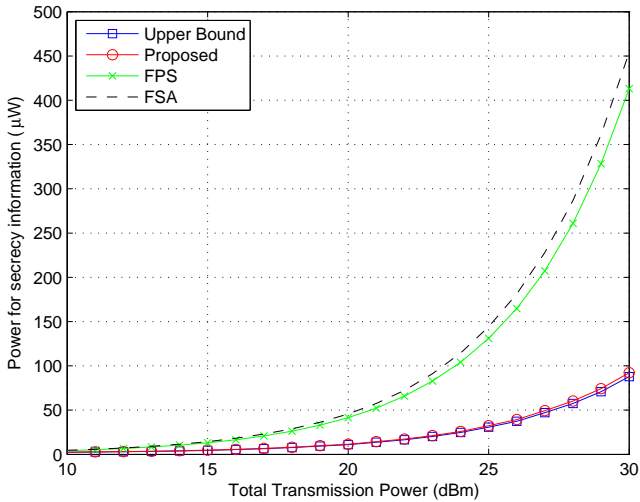}
\vspace{-0.1cm}
 \caption{Total power consumption for all information receivers versus $P_{t}$ at $\bar{C}=0.4$ bit/OFDM symbol.} \label{fig:f3}
\end{centering}
\vspace{-0.3cm}
\end{figure}

Fig. \ref{fig:f3} depicts the overall power consumption for information receivers versus the maximum transmit power $P_{t}$ under $\bar{C}=0.4$ bit/OFDM symbol.
Both algorithms consume much less power than FPS and FSA schemes.  FPS and FSA consume $300\%$ power more than the proposed algorithm.
 Moreover, the proposed algorithm performs closely to the upper bound.

\begin{figure}[t]
\begin{centering}
\includegraphics[scale=.65]{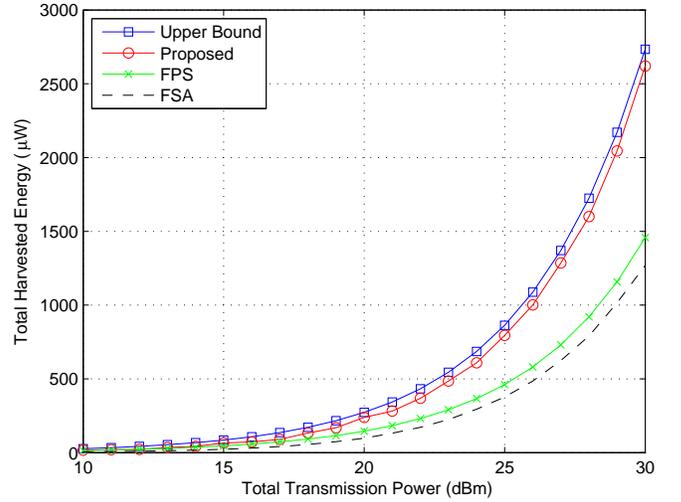}
\vspace{-0.1cm}
 \caption{Achievable $E_{sum}$ versus $P_{t}$ at $\bar{C}=0.4$ bit/OFDM symbol.} \label{fig:f4}
\end{centering}
\vspace{-0.3cm}
\end{figure}

Finally, we demonstrate the relation between $E_{sum}$ and total transmit power $P_{t}$ under $\bar{C}=0.4$ bit/OFDM symbol in Fig. \ref{fig:f4}. We can conclude that  the proposed algorithm is also very close to the performance upper bound.
At low SNR region ($P_{t}<12$dBm), all  schemes perform close to each other. As $P_{t}$ becomes larger, the
proposed algorithm only incurs a marginal performances loss, and both significantly outperform the two benchmarks.


\section{Conclusions}
    This study investigated the joint subcarrier allocation policy and power splitting ratio selection for downlink security OFDMA broadband networks. We formulated the problem to maximize the harvested power while satisfying the secrecy rate requirements of all users. Simulation results show that the proposed algorithm performs closely to upper bound and significantly outperforms the conventional methods.

\bibliographystyle{IEEEtran}
\bibliography{IEEEabrv,OFDMA}

\end{document}